\documentclass[aps,twocolumn,amsmath,amssymb,showpacs,prl]{revtex4}
\usepackage{epsf}

\newcommand{\hop}{\tau }

\renewcommand{\vec}[1]{{\bf #1}}

\newcommand{\ki}{\hat{\vec k}}

\newcommand{\vR}{\vec{R}}

\begin{document}
\title{Theory of Half Metal-Superconductor Heterostructures}
\author{M. Eschrig$^1$, J. Kopu$^1$, J. C. Cuevas$^1$, 
and Gerd Sch{\"o}n$^{1,2}$}
\affiliation{$^1$Institut f{\"u}r Theoretische Festk{\"o}rperphysik,
Universit{\"a}t Karlsruhe, 76128 Karlsruhe, Germany
\\
$^2$Forschungszentrum Karlsruhe,
Institut f{\"u}r Nanotechnologie, 76021 Karlsruhe, Germany
}
\date{June 17, 2002}
\begin{abstract}
We investigate the Josephson coupling between two singlet superconductors
separated by a half-metallic magnet. The mechanism behind the
coupling is provided by the rotation of the quasiparticle spin in
the superconductor during reflection events at the interface with
the half metal. Spin rotation induces triplet correlations in the
superconductor which, in the presence of surface spin-flip scattering,
result in an indirect Josephson effect between
the superconductors. We present a theory appropriate for studying
this phenomenon and calculate physical properties for a
superconductor/half metal/superconductor (S/HM/S) heterostructure.
\end{abstract}
\pacs{PACS numbers: 74.50.+r,73.40.-c,73.63.-b,74.80.Dm}
\maketitle 
{\it Introduction:} 
The interplay between
superconductivity and spin-polarized materials has potential
applications in the emerging field of spin electronics. For this
purpose, a high degree of spin polarization of the materials in
contact with superconducting regions is desirable.
The recently discovered half metals are ideal materials in this respect
\cite{Pickett01}. In half metals electronic bands exhibit insulating
behavior for one spin direction and metallic behavior for the other.
They are thus completely spin-polarized magnets. 
Half-metallic
behavior has been found experimentally in the manganese perovskite
La$_{0.7}$Sr$_{0.3}$MnO$_3$ \cite{Park98,Soulen98} and in CrO$_2$ \cite{Ji01}. 
The perovskite is particularly interesting because of its ability
to form high-quality heterostructures with high-$T_c$
cuprate superconductors \cite{Fu}.

The superconducting proximity effect in spin-polarized materials
has attracted considerable attention recently
in the context of superconductor/ferromagnet heterostructures
\cite{Jong95,Bergeret01,Giroud98,Petrashov99,Aprili,Hernando}.
The singlet pairing amplitude shows oscillations
with a wave vector matching the spin splitting of the Fermi
wave vectors in the ferromagnet \cite{Aprili,Buzdin}.
The magnitude
of this proximity effect decreases with increasing spin polarization.
In the extreme case of a completely spin-polarized 
material the singlet proximity effect is absent.
Consequently, the Josephson current between two singlet superconductors 
separated by a half metal is expected to be exactly zero.
In this Letter we show that this is not necessarily 
the case.
We propose a mechanism which leads to a nonvanishing S/HM/S Josephson effect.

The indirect Josephson effect requires the interplay of two
separate interface effects: spin mixing (or spin rotation) and
spin-flip scattering. The former, represented by the phase difference
$\theta$ between waves of opposite spin orientations reflected from a
spin-active interface, introduces triplet correlations at the
superconducting side of the S/HM boundary. The latter mediates these
correlations to the half-metallic side. 
To illustrate the spin-mixing effect, consider the reflection of
two quasiparticles, $\mbox{$\vert \! \! \uparrow \rangle_{{k} }$}$
and $\mbox{$\vert \! \! \downarrow \rangle_{{k} } $}$,
from a half-metallic material (which defines the spin quantization axis).
The reflected amplitudes for opposite spins differ in phase,
$\mbox{$\vert \! \! \uparrow \rangle_{-{k} }$}=
\mbox{$e^{i\theta /2} \vert \! \! \uparrow \rangle_{{k} }$} $,
$\mbox{$\vert \! \! \downarrow \rangle_{-{k} } $}
= \mbox{$e^{-i\theta /2} \vert \! \! \downarrow \rangle_{{k} }$} $
\cite{Tokuyasu88}. 
In a superconductor,
incoming quasiparticles (${k}$) near the interface
form pairs with outgoing quasiparticles ($-{k}$).
As
$\mbox{$\vert \! \! \uparrow \rangle_{{k}} $}
\mbox{$\vert \! \! \downarrow \rangle_{-{k}} $}- 
\mbox{$\vert \! \! \downarrow \rangle_{{k}}$}
\mbox{$\vert \! \! \uparrow \rangle_{-{k}}$} $
transforms under reflection into 
$e^{i\theta } 
\mbox{$\vert \! \! \uparrow  \rangle_{{k}}$}
\mbox{$\vert \! \! \downarrow \rangle_{-{k}}$}
- e^{-i\theta } 
\mbox{$\vert \! \! \downarrow \rangle_{{k}}$}
\mbox{$\vert \! \! \uparrow \rangle_{-{k}} $}$,
pairing states near such interfaces are singlet-triplet mixtures.
This property of spin mixing is intrinsic to any spin-active interface.
If, additionally, spin-flip scattering is present at the S/HM 
interface, the resulting triplet amplitudes induce equal-spin pairing correlations in
the half metal, leading to an S/HM/S Josephson effect.
Spin-flip scattering is expected to be enhanced 
{\it e.g.} due to local variations of the spin quantization axis
\cite{Bergeret01}, or diffusion of magnetic moments.
The importance of these processes was pointed out by
recent experiments \cite{Kreuzer02}.

The indirect proximity effect introduced 
above can also be relevant for strong ferromagnets. In the conventional 
description,
the dispersions for spin-up and spin-down bands in ferromagnets are assumed 
to be identical apart from an energy splitting, 
given by an effective exchange field $h$ \cite{Jong95,Bergeret01}.
The range of the spin-singlet proximity effect
is drastically reduced by a strong exchange field.
In contrast, no such suppression occurs in the case of
the indirect proximity effect.

{\it Theory:}
Our treatment is based on the quasiclassical theory of superconductivity
\cite{eilenberger68}. This theory is formulated
in terms of Green's functions (propagators) which are matrices in 
Nambu-Gor'kov particle-hole space and in spin space.
The quasiclassical propagator, $\hat g (\ki ,\vR ,\epsilon )$ depends on 
energy $\epsilon $, position $\vec{R}$, and the direction $\ki $
of the momentum on the Fermi surface.
Its particle-hole diagonal and off-diagonal elements are denoted
by spin matrices $g$ and $f$.
The quasiclassical propagator 
satisfies the Eilenberger equation \cite{eilenberger68}
\begin{equation}
\label{eq1}
\left[\epsilon \hat \tau_3 - 
\hat \Delta , \hat g\right] + i \vec{v}_f \cdot \nabla_{\vec{R}} \hat g =0 
\end{equation}
with the Fermi velocity, $\vec{v}_f(\ki )$, and the singlet order parameter $\hat \Delta(\vR)$.
It is essential for our
purpose to determine the spatial variation of the
order parameter near the interface region in accordance with the
triplet correlations, which decay into the superconducting region on the
coherence length scale.
In order to ensure current conservation in the whole system
we obtain the spatial variation of $\Delta(\vR)$ self-consistently,
\begin{eqnarray}
\Delta (\vR ) = \lambda \int^{\epsilon_c}_{-\epsilon_c} 
\frac{d\epsilon }{2\pi i } \;
\langle f(\ki,\vR,\epsilon ) \rangle_{\ki} \; \tanh\left(\frac{\epsilon }{2T}\right).
\end{eqnarray}
The coupling constant $\lambda$ and the cut-off energy $\epsilon_c$ 
are eliminated in favor of the transition temperature $T_c$
in the usual manner.
The quasiclassical Green's functions are normalized 
according to
$\hat g^2 =-\pi^2 \hat 1$ \cite{eilenberger68}.

\begin{figure}
\begin{center}
\begin{minipage}{3.4in}
\epsfxsize=2.6in
\epsfbox{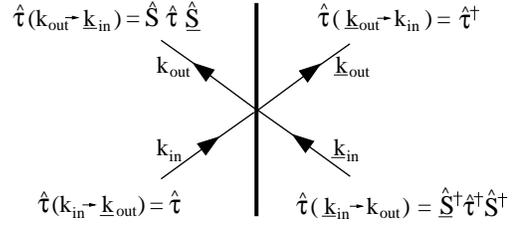}
\end{minipage}
\end{center}
\caption{ \label{hopping}
Scattering geometry illustrating the scattering channels and the
corresponding transfer amplitudes for the model discussed in the text. 
}
\end{figure}

{\it Boundary conditions:}
A standard method to treat boundary conditions for spin active
interfaces is a scattering matrix
formulation \cite{Millis88,Jong95,Fogelstrom00}.
However, for the present problem, where the number of spin channels on
one side of the interface differs from that on the other,
it would be necessary to use the formulation
by Millis {\it et al.} \cite{Millis88} which is rather involved.
For this reason we proceed with an alternative but equivalent approach
\cite{Cuevas01}.
It allows us
to derive explicit quasiclassical boundary conditions in terms of an
auxiliary Green's function,
$\hat g^0$, which solves the boundary condition for an impenetrable
interface and is easy to obtain. 
The impenetrable interface is characterized by two surface scattering
matrices, $\hat S$ and $\underline {\hat S}$, on either side of the
interface. The resulting propagators on the two sides are denoted by
$\hat g^0$ and $ \underline {\hat g}^0$, respectively. 
At the boundary, incoming 
propagators, $\hat g^0_{in}$, are connected with outgoing 
ones, $\hat g^0_{out}$, via the surface scattering matrices 
by $\hat g^{0}_{out}=\hat S \hat g^{0}_{in} \hat S^{\dagger }$
\cite{Tokuyasu88}.
Particle conservation requires unitarity, $\hat S^\dagger=\hat S^{-1}$.
We include the transmission processes through the interface via an
effective hopping amplitude $\hat \hop $
in a $t$-matrix approximation. 
We assume
translational invariance in the plane of the interface. 
The quasiclassical hopping amplitudes from left to 
right differ in general for incoming and outgoing quasiparticles. 
However, the requirement of current conservation leads to
relations between these elements as shown in Fig. \ref{hopping}.

The quasiclassical $t$-matrix equations read
\newcounter{saveeqn}%
\newcommand{\alpheqn}{\setcounter{saveeqn}{\value{equation}}%
\stepcounter{saveeqn}\setcounter{equation}{0}%
\renewcommand{\theequation}
        {\mbox{\arabic{saveeqn}\alph{equation}}}}%
\newcommand{\reseteqn}{\setcounter{equation}{\value{saveeqn}}%
\renewcommand{\theequation}{\arabic{equation}}}%
\alpheqn
\begin{eqnarray}
\label{tmatrix}
\hat t_{in}&=& \hat \hop \underline {\hat g}^{0}_{out} \hat \hop^{\dagger }
\big( \hat 1 + \hat g^{0}_{in} \hat t_{in} \big)
\; , \quad 
\hat t_{out}= \hat S \hat t_{in}  \hat S^{\dagger } 
, \qquad \; \\
\underline{\hat t}_{out}&=& \hat \hop^{\dagger } {\hat g}^{0}_{in} \hat \hop
\big( \hat 1 + \underline{\hat g}^{0}_{out} \underline{\hat t}_{out} \big)
\; , \quad 
\underline{\hat t}_{in}= \underline{\hat S}^{\dagger } 
\underline{\hat t}_{out}  \underline{\hat S}
\; .  \qquad \;
\end{eqnarray}
\reseteqn
On each side of the interface, the $t$ matrix describes the
modifications of the quasiclassical propagators due to 
virtual hopping processes to the opposite side.
Finally, we express the full propagator in terms of the 
decoupled solution $\hat g^0$,
leading to the boundary
conditions for incoming and outgoing propagators,
\alpheqn
\begin{eqnarray}
\hat g_{in}&=&\hat g^{0}_{in} +
\Big\{\hat g^{0}_{in} + i\pi\hat 1 \Big\} \; 
\hat t_{in} \; \Big\{\hat g^{0}_{in} - i\pi\hat 1 \Big\}, \qquad
\\
\hat g_{out}&=&\hat g^{0}_{out} +
\Big\{\hat g^{0}_{out} - i\pi\hat 1 \Big\} \;
 \hat t_{out} \; \Big\{\hat g^{0}_{out} + i\pi\hat 1 \Big\}, \qquad
\label{eq7}
\end{eqnarray}
\reseteqn
and similarly for $\underline{\hat g}_{in}$ and
$\underline{\hat g}_{out}$ \cite{CurrCons,Keldysh}.
In the appropriate limiting cases these
boundary conditions
reduce to those published previously \cite{Zaitsev84,Millis88,Cuevas01,Fogelstrom00,Eschrig00}.

For reference, we also present the corresponding 
full scattering matrix which would enter
the boundary conditions of Ref. \cite{Millis88}.
Without loss of generality 
it can be written in the form
\begin{equation}
\label{scatt}
\hat {\bf S}= 
\left(
\begin{array}{cc}
\hat S_{11} & \hat S_{12} \\ \hat S_{21} & \hat S_{22}
\end{array}
\right)
=
\left(
\begin{array}{cc}
\hat S & 0 \\ 0 & {\hat 1}
\end{array}
\right)
\left(
\begin{array}{cc}
\hat r & \hat d \\
\hat d^{\dagger} & 
-\underline{\hat r} 
\end{array}
\right)
\left(
\begin{array}{cc}
\hat 1 & 0 \\ 0 & \underline{\hat S}
\end{array}
\right)
\end{equation}
with the transmission matrix
$\hat d= (1+\pi^2\hat\hop \hat\hop^{\dagger })^{-1} 2\pi \hat\hop $, and
the reflection matrices on either side of the interface,
$\hat r= (1+\pi^2\hat\hop \hat\hop^{\dagger })^{-1}
(1-\pi^2\hat\hop \hat\hop^{\dagger })$,
$\underline{\hat r}=
(1+\pi^2\hat\hop^{\dagger }\hat\hop )^{-1}
(1-\pi^2\hat\hop^{\dagger }\hat\hop )$.

The particle-hole structures of the 
surface scattering matrix 
and the 
hopping amplitude 
are
$\hat S = $diag$[S,\tilde{S}]$
and
$\hat \hop = $diag$[\hop ,\tilde S^{\dagger } 
\hop^{\ast } \underline{\tilde{S}}^{\dagger }]$.
The above equations are for general spin structures. In the following,
$\tau $ is a 2x1 spin matrix, $S$ a 
2x2 spin matrix, and $\underline S$ a 
spin scalar.

{\it S/HM/S structure:} We study a 
heterostructure consisting of a half metal, $-L_H\!<\!x\!<\!L_H$,
between two superconductors, 
$-L\! <\! x\! <\! -L_H$ and $L_H\! <\! x\! <\! L$.
We investigate the equilibrium supercurrent
due to a phase difference $\phi $ between the superconductors, 
$\Delta (L) = \Delta (-L) e^{i\phi }$. 

As mentioned above, band splitting in the interface region
results in a relative spin phase for quasiparticles with
spin along the quantization axis of the half metal
(for quasiparticles with spin in the perpendicular plane 
the corresponding effect is a spin rotation around the quantization axis)
\cite{Tokuyasu88}. 
This effect can be described by a scattering matrix $\hat S= \exp(i\theta
\sigma_z/2) \hat 1 $ at the superconducting side of the interface,
where $\theta $ defines a spin-rotation angle 
and $\sigma_z$ denotes the Pauli spin matrix \cite{Tokuyasu88,Fogelstrom00}. 
Generally, the value of $\theta $ depends on the angle of impact, 
$\psi $ \cite{Tokuyasu88}
and can approach values of the order of $\pi $ for strong band 
splitting \cite{Barash02}. For definiteness, we present results for
$\theta = 0.75 \pi \; \cos \psi $.
On the half-metallic side the scattering matrix has no spin structure,
$\underline{\hat S}=\hat 1$.

The $t$-matrix equations
are parameterized by the hopping amplitude $\hat \hop $
and the scattering matrices $\hat S$, $\underline{\hat S}$, which are
the phenomenological parameters characterizing the interface in our theory. 
We use $\hop = (1+S^{\dagger})
\hop_0 \cos \psi $, where
$\hop_0 
=(\tau_{\uparrow \uparrow},\tau_{\downarrow \uparrow})^{\rm T}
$ 
is determined by the two spin scattering channels
from the superconductor to the half-metallic spin-up band.
With this choice the spin rotation during transmission is half of the spin
rotation during reflection.
The $\cos \psi$ factor accounts for the reduced transmission 
at large impact angles.
We present results for $\hop_{\downarrow\uparrow}/ \hop_{\uparrow\uparrow}=
0.7$ and $0.1$, $2\pi \hop_{\uparrow\uparrow}=1.0$,
$2L_H=3 \xi_0$ 
(with the coherence length $\xi_0= v_f/2\pi T_c$), 
$L \gg L_H$, 
and cylindrical Fermi surfaces 
(calculations using spherical Fermi surfaces lead to similar results).
We iterate Eqs.~(1) and (2) 
until self-consistency is achieved, with the boundary 
conditions (3) and (4) at the two interfaces.
All our calculations are in the clean limit.
\begin{figure}
\begin{minipage}{3.4in}
\epsfxsize=3.0in
\epsfbox{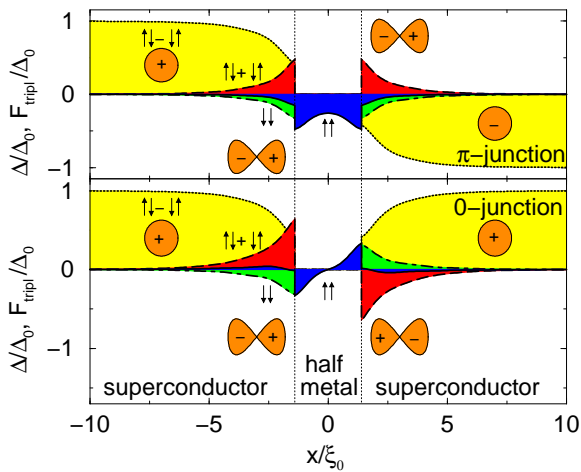}
\end{minipage}
\caption{ \label{fig1}
Self-consistent order parameter and triplet correlations in an
S/HM/S heterostructure for a zero 
junction and a $\pi$ junction. The relative signs of the pairing correlations
in the $s$-wave singlet and three $p$-wave triplet channels are indicated.
A zero junction for the singlet order parameter leads to a 
relative phase difference of $\pi$ for the triplet correlations, 
and vice versa. The calculations are for temperature $T=0.05 T_c$,
and for $\hop_{\downarrow \uparrow}/\hop_{\uparrow \uparrow}=0.7$.
}
\end{figure}

In Fig.~\ref{fig1} we present the 
spatial modulation of the singlet order parameter and the triplet
pairing correlations for an S/HM/S heterostructure.
We compare results for a zero junction ($\phi=0$) and a $\pi$
junction ($\phi=\pi)$.
The spin-rotation effect at the superconducting side of the interface
leads to local triplet correlations in the superconductor of the form
$f_{\uparrow\downarrow}+f_{\downarrow\uparrow}$. 
We quantify the triplet pairing correlations by the
integral
\begin{eqnarray}
F_{tripl} (x ) = \int^{\epsilon_c}_{-\epsilon_c} \frac{d\epsilon }{2\pi i} 
\; \langle \eta (\ki ) f(\ki,x,\epsilon ) \rangle_{\ki} \; 
\tanh\left(\frac{\epsilon }{2T}\right),
\end{eqnarray}
where $\eta (\ki)$ projects out the $p$-wave pairing amplitude, and is
equal to the cosine of the angle between $\ki $ and the surface
normal. 
Spin-flip scattering induces a $F_{\uparrow\uparrow}$ amplitude in
the half metal, and leads to
both $F_{\uparrow\uparrow}$ and $F_{\downarrow\downarrow}$ amplitudes in
the superconductor.
The correlations are shown in Fig.~\ref{fig1} for all three
spin-triplet channels. Triplet correlations extend into the
superconductor up to a few coherence lengths from the interface,
leading to a suppression of the singlet order parameter near the
interface. We also show schematically the $s$ and $p$ orbitals for a
zero junction and a $\pi$ junction. The alignment of the $p$ orbitals
is determined by the direction of the surface normal. As a
consequence, the relative sign between the $p$ orbitals is opposite to
that of the $s$ orbitals. As will be shown below, this leads to a
reversal of the current 
direction from that expected for a
superconductor/normal metal/superconductor junction.

\begin{figure}
\begin{minipage}{3.4in}
\epsfxsize=3.0in
\epsfbox{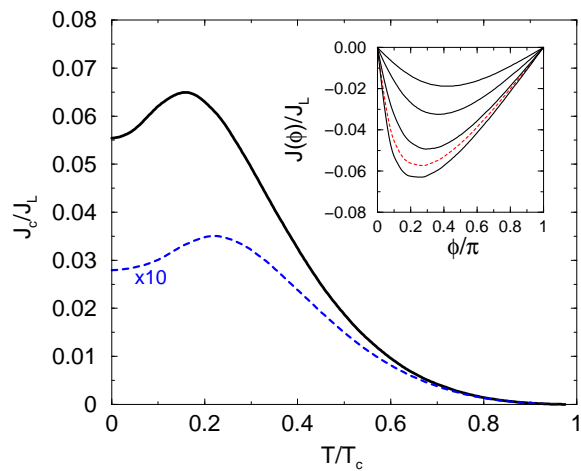}
\end{minipage}
\caption{ \label{fig3}
  Critical Josephson current density as a function of temperature for an
  S/HM/S heterostructure. The two curves are for 
  $\hop_{\downarrow \uparrow}/\hop_{\uparrow \uparrow}=0.1$ (dashed) and 
  $\hop_{\downarrow \uparrow}/\hop_{\uparrow \uparrow}=0.7$ (full lines).
  The inset shows the current-phase relationships 
  for $\hop_{\downarrow \uparrow}/\hop_{\uparrow \uparrow}=0.7$ 
  for temperatures
  $T/T_c=0.05$ (dashed), $0.2,0.3,0.4,0.5$ (full lines from
  bottom to top). The unit is the Landau critical current density
  $J_L=ev_f N_f\Delta_0$, with the zero temperature bulk superconducting
  gap $\Delta_0=1.76 T_c$.  }
\end{figure}

We now turn to the half-metallic region
in Fig. \ref{fig1}. The spatial distribution of the
proximity-induced $F_{\uparrow\uparrow}$ amplitude 
shows a sign change at $x=0$ in the case of zero junction, 
but not for a $\pi$ junction. 
As a result, the $\pi $ junction is expected to be 
more stable than the zero junction. 
Indeed, our numerical calculations show 
that the $\pi $ junction corresponds to the free-energy minimum
for all temperatures.  
The equal-spin correlations decay slowly into the half metal, {\it e.g.} 
$F_{\uparrow\uparrow}(x=0)\propto 1/L_H$ in the $\pi$ junction. This 
behavior is similar to that observed in normal metal/superconductor
structures.

In Fig. \ref{fig3} we show the Josephson 
critical current as a function of
temperature. 
The current density,
\begin{eqnarray}
\vec{J} 
= \int^{\infty}_{-\infty} d\epsilon 
\; \langle e \; \vec{v}_f(\ki ) \; N_\uparrow(\ki,\epsilon ) 
\rangle_{\ki} \; n_f(\epsilon ),
\label{current}
\end{eqnarray}
is expressed in terms of
the angle-resolved density of states at the Fermi surface
in the half metal,
$N_\uparrow=-N_f \; {\rm Im} (g_{\uparrow\uparrow})/\pi $, 
the electronic charge $e$, 
and the Fermi distribution function $n_f$. 
In the inset of Fig. \ref{fig3} we show the current-phase relationship for
different temperatures. 
The current is negative for a positive 
phase difference $\phi $. 
For each temperature we determine the critical current from
the maximum current magnitude in the current-phase relationship.
The critical current has a $(1-T/T_c)^2$ 
dependence near $T_c$. This is a consequence of the fact that the order
parameter at the interface varies linearly with $1-T/T_c$, in contrast
to the bulk 
$(1-T/T_c)^{1/2}$ 
\begin{figure}
\begin{minipage}{3.4in}
\centerline{
\epsfxsize=1.5in{\epsfbox{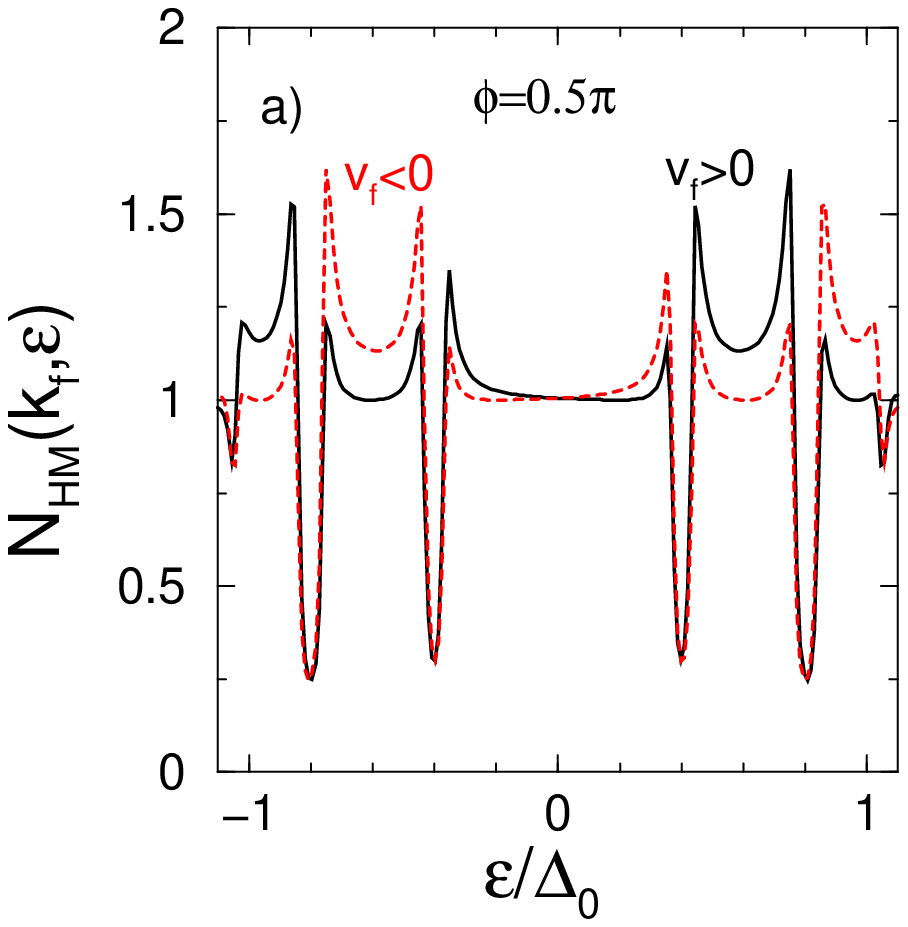}}
\epsfxsize=1.5in{\epsfbox{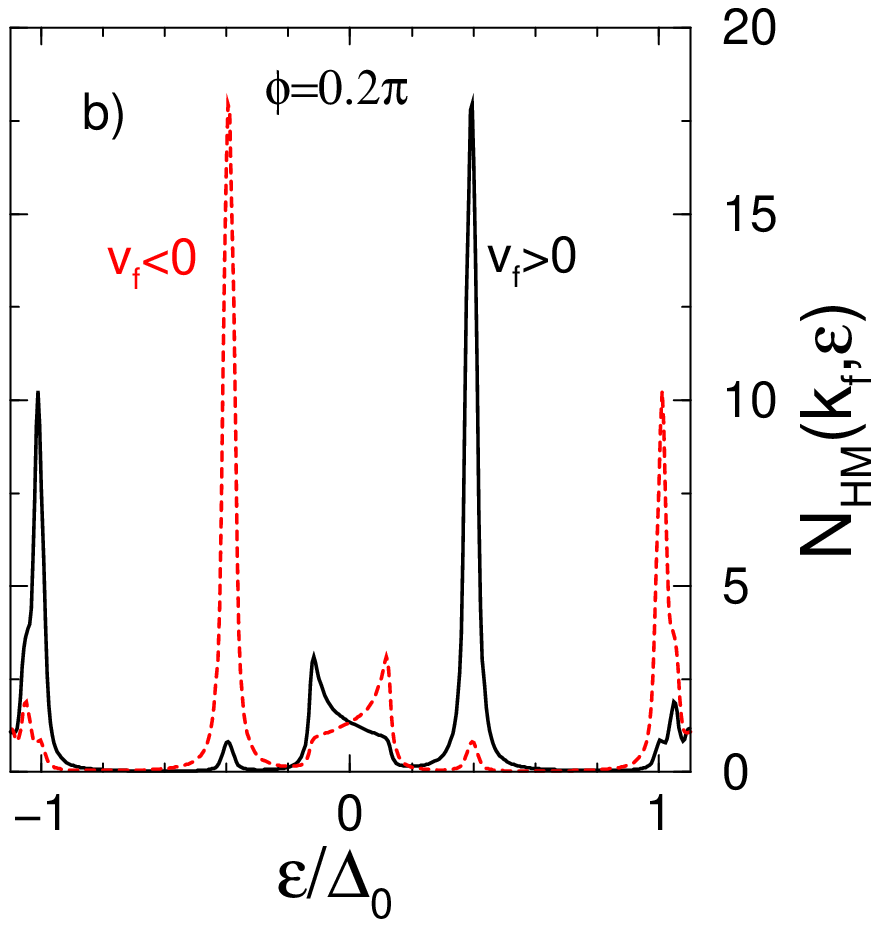}}
}
\end{minipage}
\caption{ \label{fig2}
Density of states at $T=0.05T_c$
for quasiparticles 
with normal impact at the half-metallic side of the left interface
($x=-L_H $), for a) 
spin flip rate $\hop_{\downarrow \uparrow}/\hop_{\uparrow \uparrow}=0.1$ and
phase difference $\phi =0.5 \pi$, and b)
$\hop_{\downarrow \uparrow}/\hop_{\uparrow \uparrow}=0.7$ and $\phi =0.2 \pi$.
The corresponding Josephson currents are close to the critical values.
Shown are both states carrying current in positive (full lines)
and negative directions (dashed lines). 
}
\end{figure}
\noindent 
behavior. 
At low temperatures the critical current passes through a maximum and
then decreases again. This anomaly is due to the interplay between
current-carrying states, as we proceed to explain.

We discuss the different contributions to the Josephson
current coming from the spectral features in the momentum-resolved density of
states $N_\uparrow$ in the half metal.
The total current through the interface is dominated by quasiparticle
trajectories parallel to the surface normal.
In Fig. \ref{fig2} we compare the spectrum of such quasiparticles
for incoming and outgoing momenta at the half-metallic
side of the left interface.
We present results for 
$\hop_{\downarrow \uparrow}/\hop_{\uparrow \uparrow }=0.1$ and
$\hop_{\downarrow \uparrow}/\hop_{\uparrow \uparrow }=0.7$. 
In both cases there is a continuum around the chemical potential ($\epsilon=0$).
On either side of this continuum there is a gap, followed by either additional
continuum branches, or by Andreev bound states. The Andreev bound states in
Fig. \ref{fig2}b are closely related to the surface Andreev states discussed
in Refs. \onlinecite{Fogelstrom00,Barash02}.
According to Eq.~(\ref{current}), the current is obtained by
multiplying the curves in Fig. \ref{fig2} with the Fermi function. 
At not too low temperatures the Josephson current is dominated by
the negative-energy features below the continuum at the 
chemical potential. 
These features carry current in {\it negative} direction, explaining the 
negative sign of the Josephson current for positive phase difference. 
Below a certain temperature, the corresponding states are fully populated,
and the temperature dependence of the Josephson current is dominated by
the low-energy continuum around the chemical potential.
The current carried by this low-energy band is {\it positive} and increases
with decreasing temperature, 
leading to the decrease of the
magnitude of the critical Josephson current at low temperatures in 
Fig.~\ref{fig3}.

{\it Conclusions:} We have presented a theory for half
metal-superconductor heterostructures and have investigated the
Josephson coupling through a half-metallic layer with a thickness of
several coherence lengths. The Josephson coupling is induced by
triplet pairing correlations in the superconductor. These triplet
correlations are coupled to the singlet superconducting order
parameter via a spin-rotation effect, which occurs when quasiparticles
in the superconductor are reflected from a spin-polarized medium. We
have performed self-consistent numerical calculations for this
problem, and found a low-temperature anomaly in the temperature
behavior of the critical Josephson current. This anomaly is a robust
feature, which is not very sensitive to parameter variations. 
We discuss the Andreev excitation spectrum in the half metallic
region, and explain the temperature variation of the Josephson current
in terms of this spectrum.

This work was supported 
by the Deutsche Forschungsgemeinschaft within
the Center for Functional Nanostructures.


\begin{thebibliography}{99}
\bibitem{Keldysh} 
Eqs. (3a)--(4b) also can be used for Keldysh-Green's functions
in non-equilibrium Keldysh formalism. Matrix products then include
time convolutions.
\end{thebibliography}
\end{document}